\documentclass{PoS}

\title{Search for high mass resonances in dilepton, dijet and diboson final states at the Tevatron}

\ShortTitle{Search for high mass resonances}

\author{\speaker{Michel Jaffr\'e}\\
        on behalf of the CDF and D0 collaborations\\
        Laboratoire de l'Acc\'el\'erateur Lin\'eaire\\
        Universit\'e Paris-Sud\\
        CNRS-IN2P3\\
        E-mail: \email{jaffre@lal.in2p3.fr}}

\abstract{At hadron colliders, new massive particles can be searched for by 
the observation of high tranverse momentum objects forming high-mass resonances.
Searches for additional massive vector bosons (W',Z'), Randall-Sundrum gravitons
and sneutrinos in R-parity violating scenarios are performed in dilepton, dijets
and diboson final states.
The most recent results from the CDF and D0 experiments at the Tevatron are presented corresponding to integrated luminosities between 1 and 4~fb$^{-1}$.
}

\FullConference{European Physical Society Europhysics Conference on High Energy Physics,
EPS-HEP 2009,\\
		 July 16 - 22 2009\\
		 Krakow, Poland}

\begin{document}

\section{Introduction}
  Are there any new gauge bosons apart from those associated with the
SU(3)xSU(2)xU(1) group which describes our standard model (SM) of particles
and fields ?
They are predicted by many theories which have been developed to answer
some questions left open by the SM.
 Neutral gauge bosons of spin 1, labelled Z', are associated with new U(1) groups.
Existence of extra spatial dimensions is a possible explanation for the large
hierarchy between the electroweak and gravity scales.
In the Randall-Sundrum model (RS), the Kaluza-Klein (KK) excitations of the graviton
can be produced as resonances at colliders.
Lastly, spin 0 resonances like a sneutrino, can be produced in supersymmetric
theories (SUSY) when the R parity is violated.
This report describes the analysis separately for each final state.  

\section{Dielectron resonance search}
\begin{figure}[hbt]
\centering
\begin{minipage}{0.90\textwidth}
  \centering
  \begin{picture}(145,40)
    \put(6,0){\hbox{
      \includegraphics[height=4.4cm,width=6.0cm]{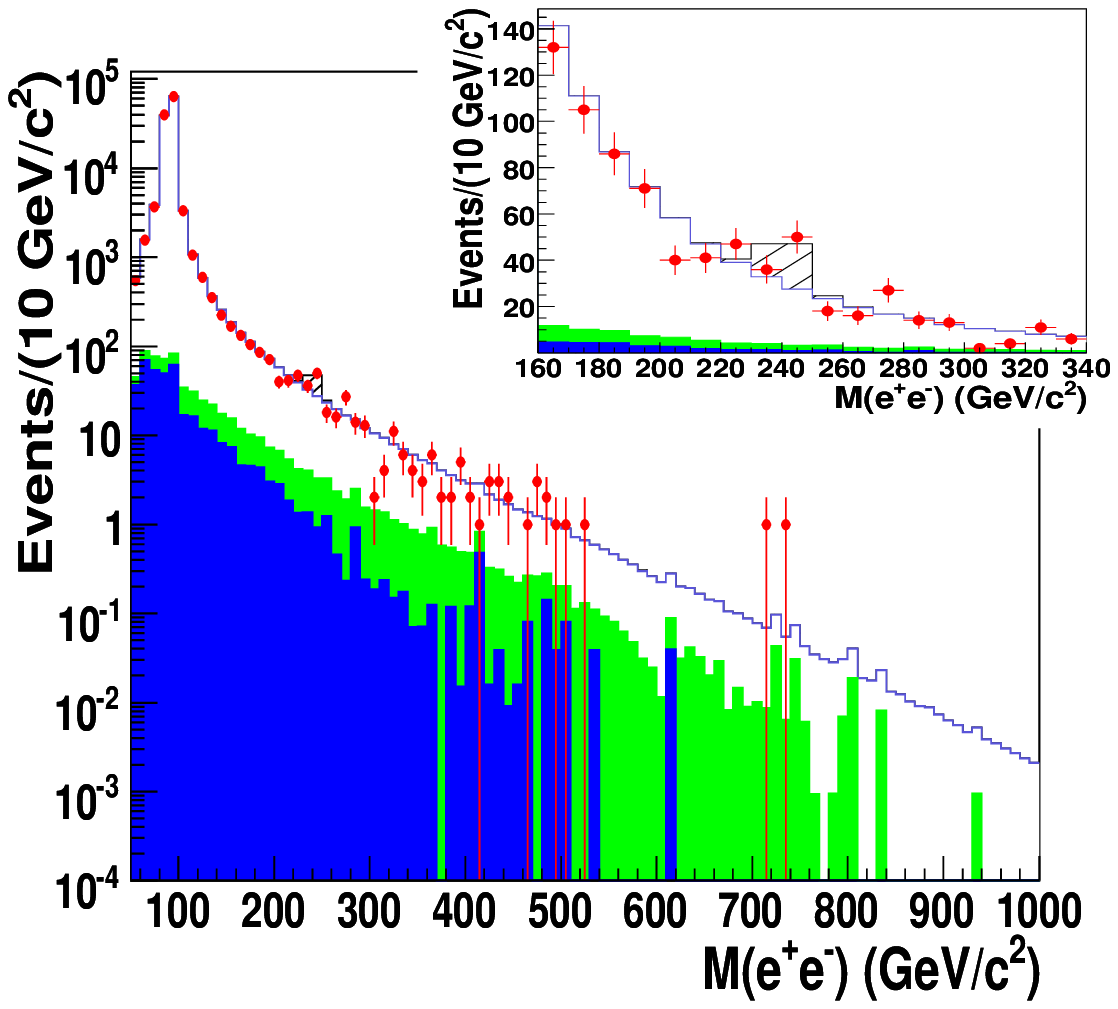}
      \includegraphics[height=4.4cm,width=6.0cm,bb=0 8 566 383]{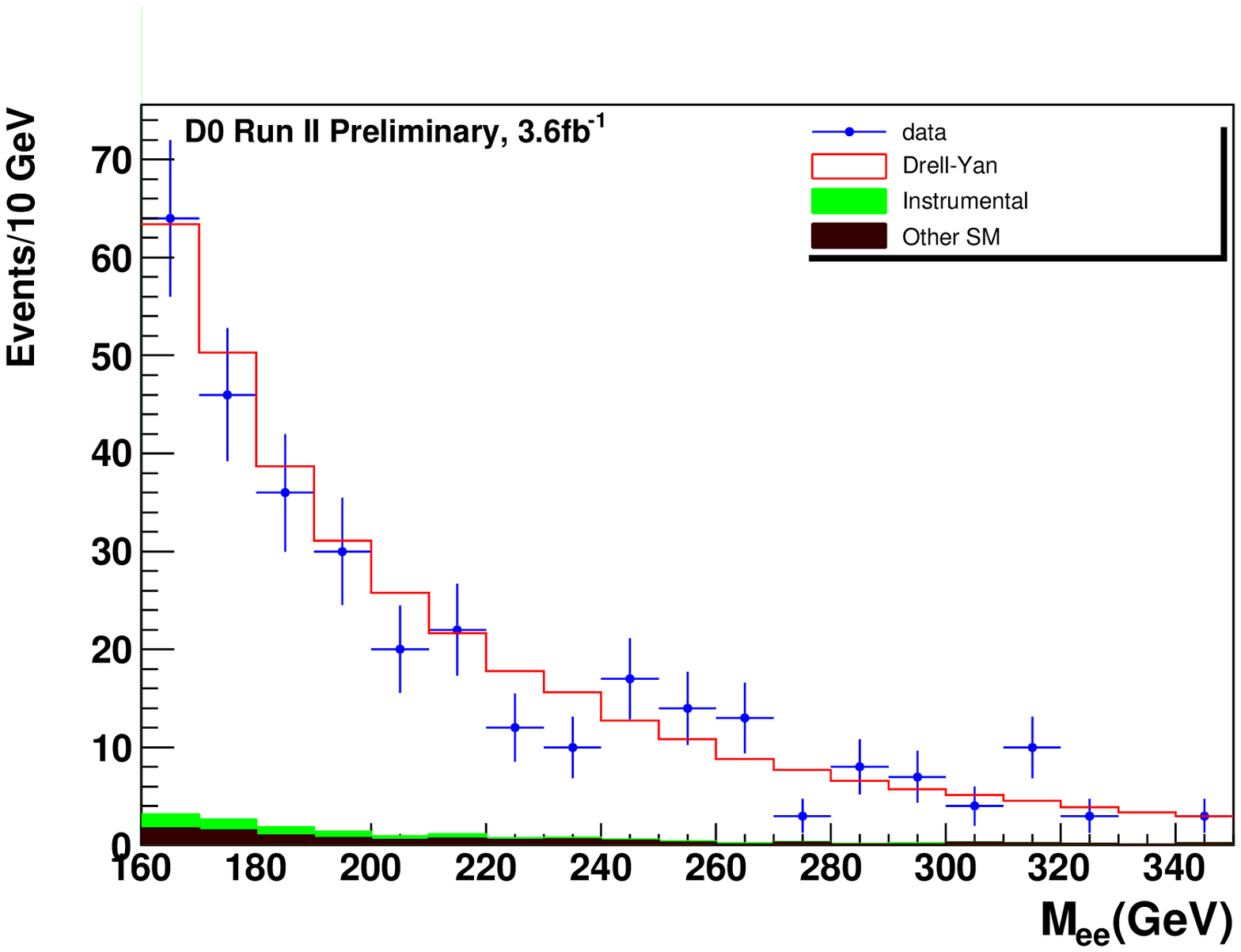}
    }}
    \put(19,37){\tiny \bf CDF RunII}
    \put(24,32){\small(a)}
    \put(85,32){\small(b)}
  \end{picture}
  \caption{\label{fig:mee}
    Invariant mass distribution of $e^+e^-$ events compared to the
    expected backgrounds for CDF (a) and D0 (b).}
\end{minipage}
\end{figure}

CDF has published~\cite{cdf-ee} an analysis based on 2.5~fb$^{-1}$ of data events
with two electrons in the final state
requiring the first electron to be central and the second to be either central
or forward.
If both electrons are central, they must have opposite charges. 
The dielectron mass distribution is shown on Fig.~\ref{fig:mee}a 
where data is compared with the expected background, essentially Drell-Yan
with small contributions from other SM sources and QCD when a jet or a photon is
misidentified as an electron.
A $2.5\sigma$ discrepancy appears around 240~GeV as better seen in the figure 
inset.
The probability to get such a fluctuation anywhere above 150~GeV is estimated to be 0.6\%.  

D0 has adapted its published analysis on the forward-backward measurement to present
the dielectron mass distribution obtained from a larger dataset of about 3.6~fb$^{-1}$~\cite{d0-ee}.
It does not support the excess of events quoted by CDF (see Fig.~\ref{fig:mee}b).
%
%

Since at higher masses  data and the SM expectation agrees,
both experiments proceed to set limits on the cross section times the branching ratio for 
signal of new physics as a function of the dielectron mass.
Although the D0 dataset is larger, its statistics suffers from very tight selection on central electrons, then only CDF limits will be reported here.

\begin{figure}[htb]
\centering
\begin{minipage}{0.90\textwidth}
  \centering
  \begin{picture}(145,40)
    \put(9,0){\hbox{
     \includegraphics[width=6cm]{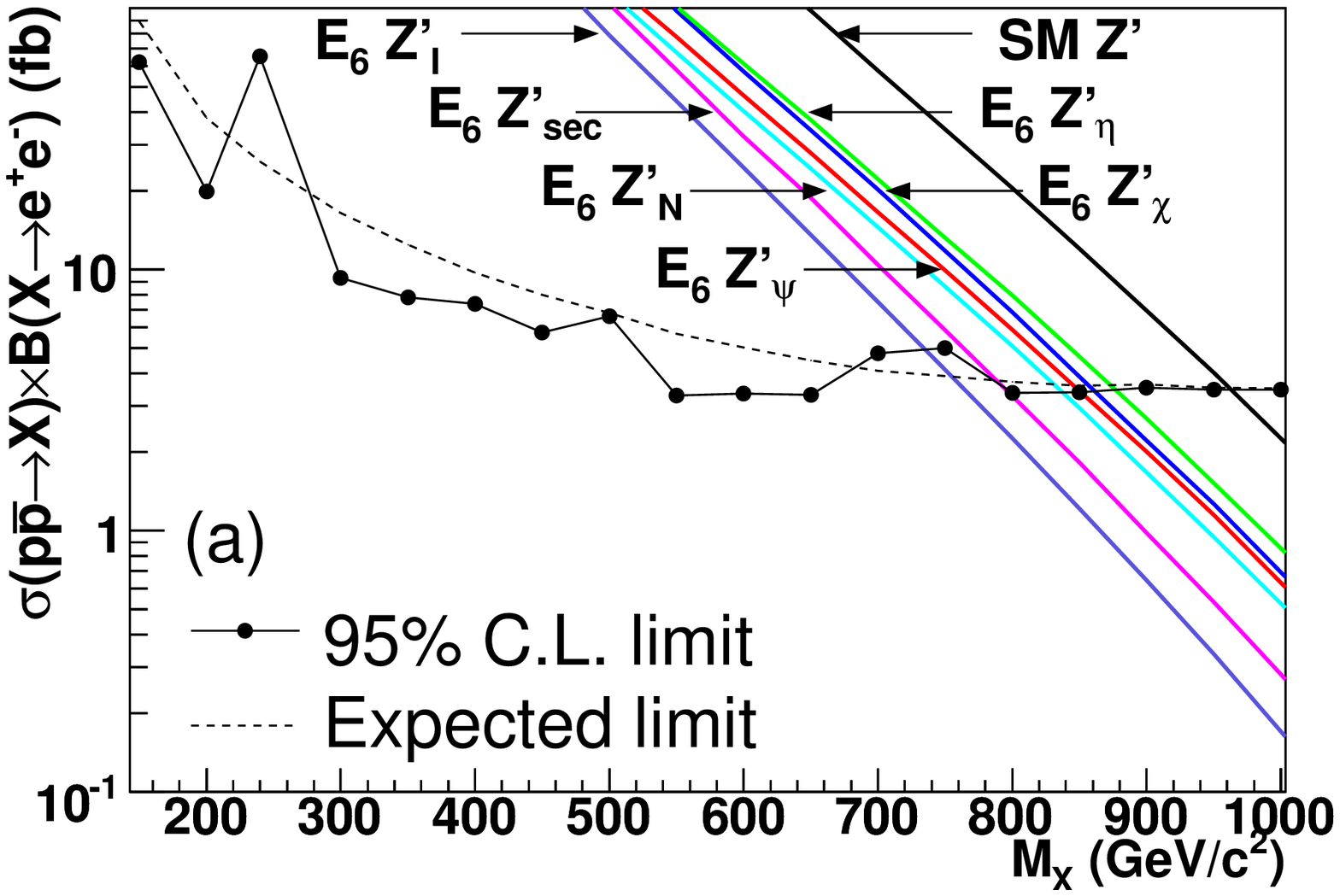}
     \includegraphics[width=6cm]{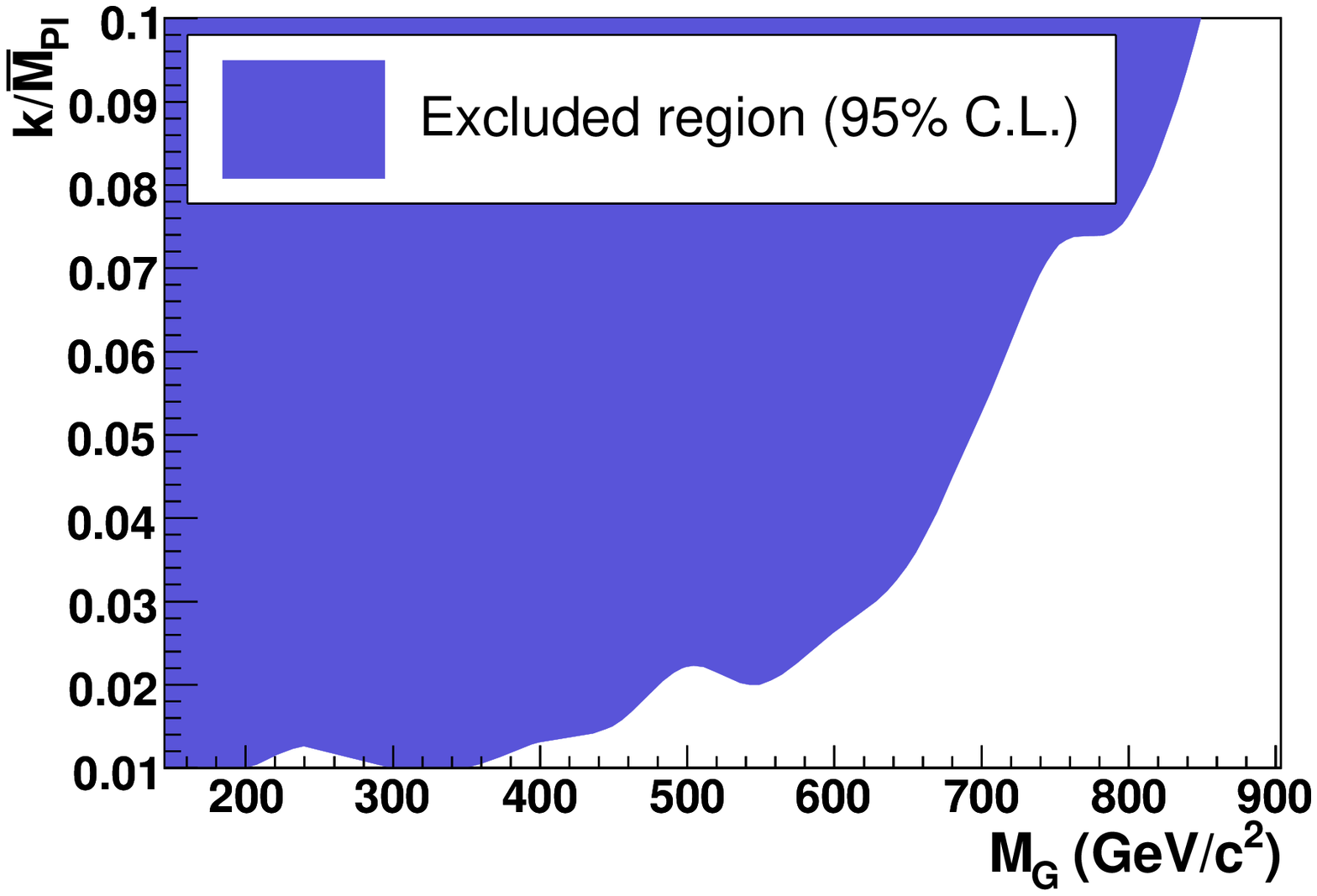}
    }}
    \put(119,13.5){\small(b)}
  \end{picture}
  \caption{(a) The upper limits on $\sigma(p\bar{p}\rightarrow
  X)\cdot\mathcal{B}(X\rightarrow e^+e^-)$ as function of the mass of an $X$
  particle at the 95\% C.L. where $X$ is a spin 1 particle.
 Also shown are the theoretical cross sections for various Z' models~\cite{zprimeModels}.
 (b) $k/\overline{M}_{Pl}$ as a function of the graviton mass in the RS model.
  level.}
  \label{fig:cdf-limits}
\end{minipage}
\end{figure}

These limits are compared to various predictions of Z' models~\cite{zprimeModels}
including the SM-like Z'.
in Fig.~\ref{fig:cdf-limits}a. A SM-like Z' is excluded below a mass of 963~GeV.

Fig.~\ref{fig:cdf-limits}b shows the excluded domain in the 2 parameter space of the
RS model: the dimensionless coupling constant $k/M_{Pl}$ whose value should lie between 0.01 and 0.1,
and  the mass of the lowest KK excited state.  
%
%
\vspace{-8mm}
\section{Dimuon resonance search}
\begin{figure}[hbt]
\centering
\begin{minipage}{0.90\textwidth}
  \centering
  \begin{picture}(145,37)
    \put(5,0){\hbox{
      \includegraphics[width=6cm] {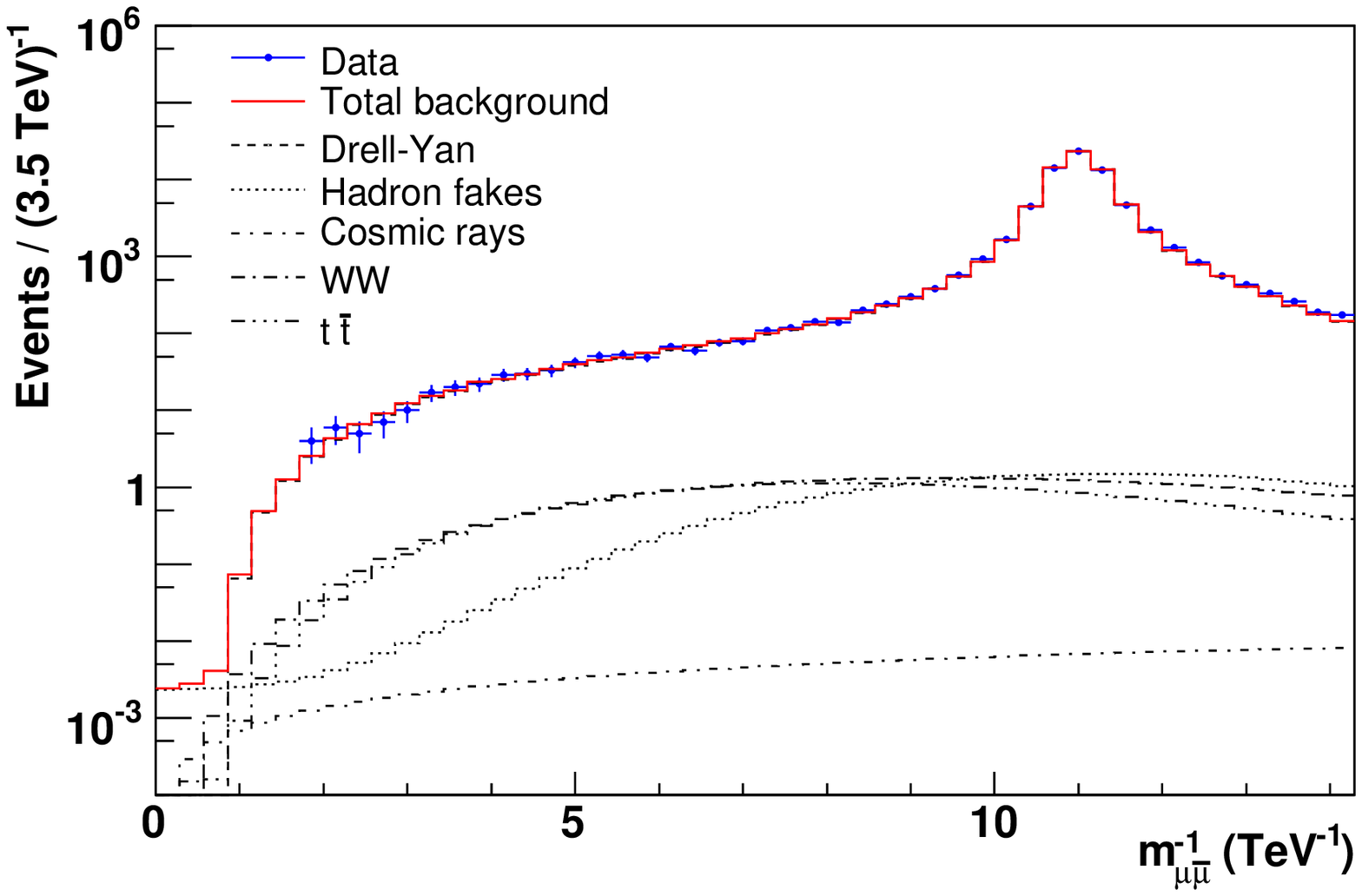}
      \includegraphics[height=3.7cm] {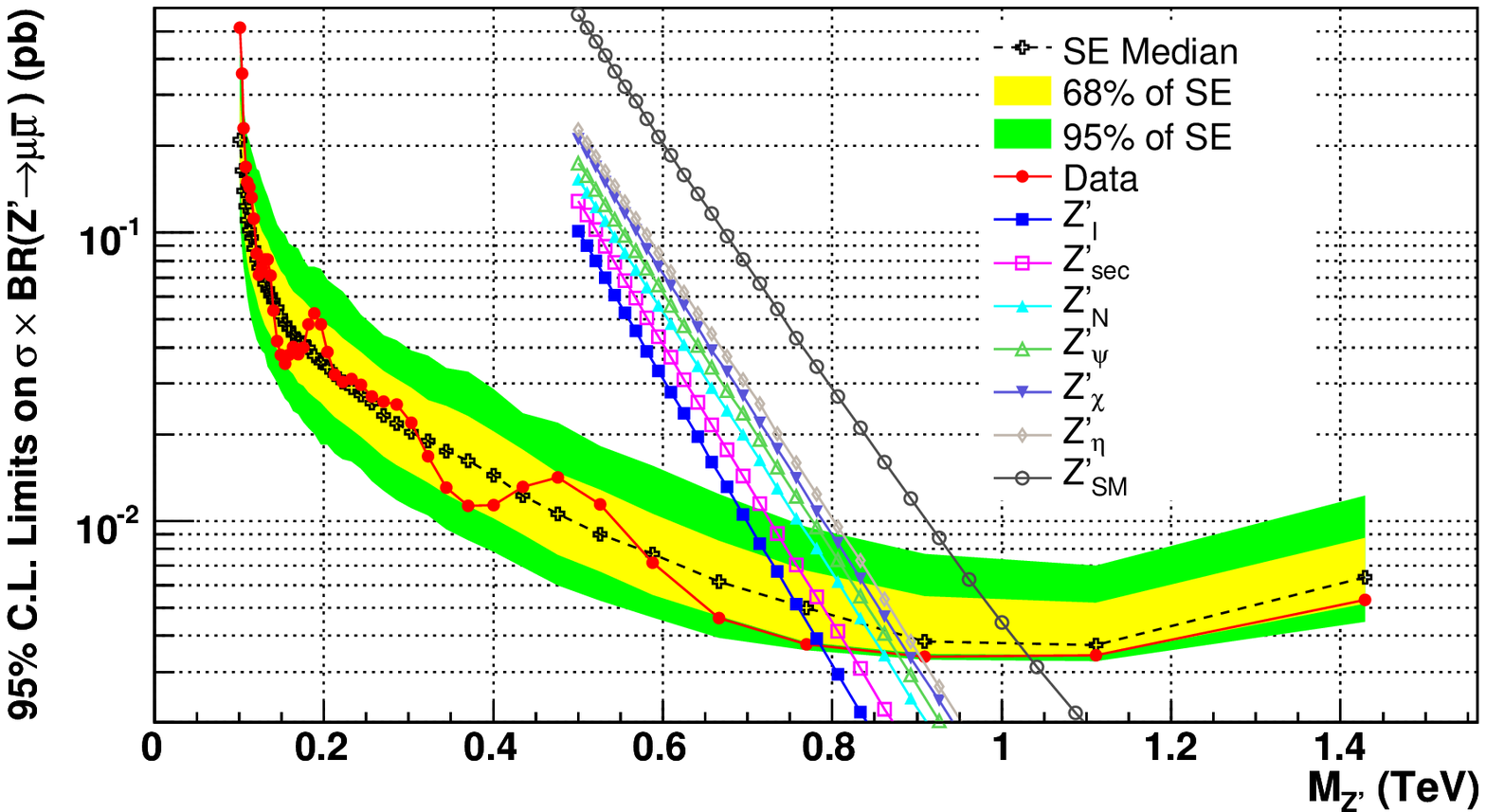}
    }}
    \put(20, 9){(a)}
    \put(80, 9){(b)}
  \end{picture}
  \caption{(a) The distribution of $m_{\mu\mu}^{-1}$ (TeV$^{-1}$) for the observed data (points),
 the individual backgrounds (dotted or dashed histograms) and the summed background (solid histogram).
(b) The 95\% C.L. upper limits on 
 $\sigma \cdot BR (Z^\prime \rightarrow \mu \bar{\mu})$ versus $M_{Z'}$.
 Also shown are the theoretical cross
 sections for various Z' models}.
  \label{fig:cdf_zprimemm_invmass_fit_data}
\end{minipage}
\end{figure}

 CDF has published a similar analysis requiring the 2 leptons to be 2 central muons of opposite charge, based on a 2.3~fb$^{-1}$ data sample~\cite{cdf-mumu}. 
For dimuon masses above 200~GeV, the resolution on the inverse of the invariant
mass ($m_{\mu\mu}^{-1}$), dominated by the track curvature resolution, is approxi\-mately constant and equal to 0.17~TeV$^{-1}$.
Fig.~\ref{fig:cdf_zprimemm_invmass_fit_data}a shows the $m_{\mu\mu}^{-1}$ distribution.
The background consists essentially of SM Drell-Yan events, with a small
contribution from $W^+W^-$ and $t\bar{t}$ SM processes, and backgrounds from muon
misidentification (cosmics, decay in flight and QCD jets).
Sum of backgrounds is normalised to data in the Z peak range between 70 and 100~GeV.
%
%

\vspace{-5mm}
A narrow resonance would appear as an excess of events in 3 adjacent bins
in this distribution.
 There is no such an excess ( in particular around 240~GeV) indicating that
data observation is consistent with SM predictions. 
CDF proceeds to set limit on the production cross section times branching
ratio for a spin 0, a spin 1 or a spin 2 particle as a function of the 
particle mass.
Fig.~\ref{fig:cdf_zprimemm_invmass_fit_data}b shows the limit for a spin 1 particle compared
with various Z' model predictions~\cite{zprimeModels}.
A SM-like Z' is excluded if its mass is below 1030~GeV.
\section{Electron muon resonance search}
D0 has searched for a $e\mu$ resonant state in a new data sample of about 3.1~fb$^{-1}$ as a followup of its published result with 1~fb$^{-1}$~\cite{d0-emu} .
The main background contributions are from $Z\rightarrow\tau\tau$, WW/WZ,
$t\bar{t}$ and W + a jet faking an electron. WW and $t\bar{t}$ processes 
are suppressed by requiring a missing transverse energy $E_T$ < 20~GeV not 
aligned or antialigned with the muon candidate (to take into account muon
 $p_T$ mismeasurement).
As there is no jet activity in signal apart from initial state QCD radiation,
events with a jet of $p_T$ > 25~GeV are rejected.

\begin{figure}[htb]
\centering
\begin{minipage}{0.90\textwidth}
  \centering
  \begin{picture}(145,40)
    \put(5,0){
      \includegraphics[width=6cm]{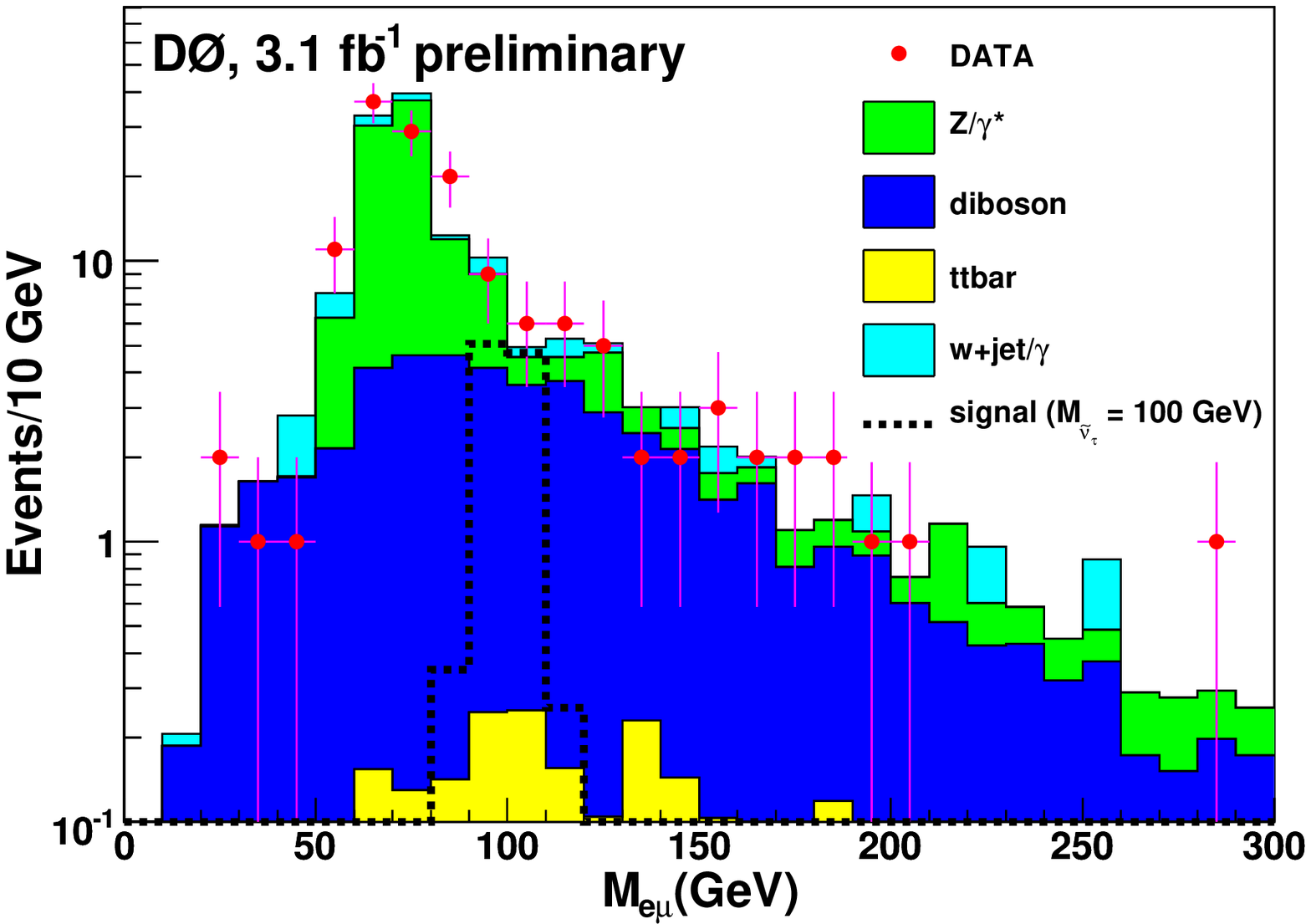}
    }
    \put(66,-3){
       \includegraphics[width=6.5cm,bb=0 0 593 400]{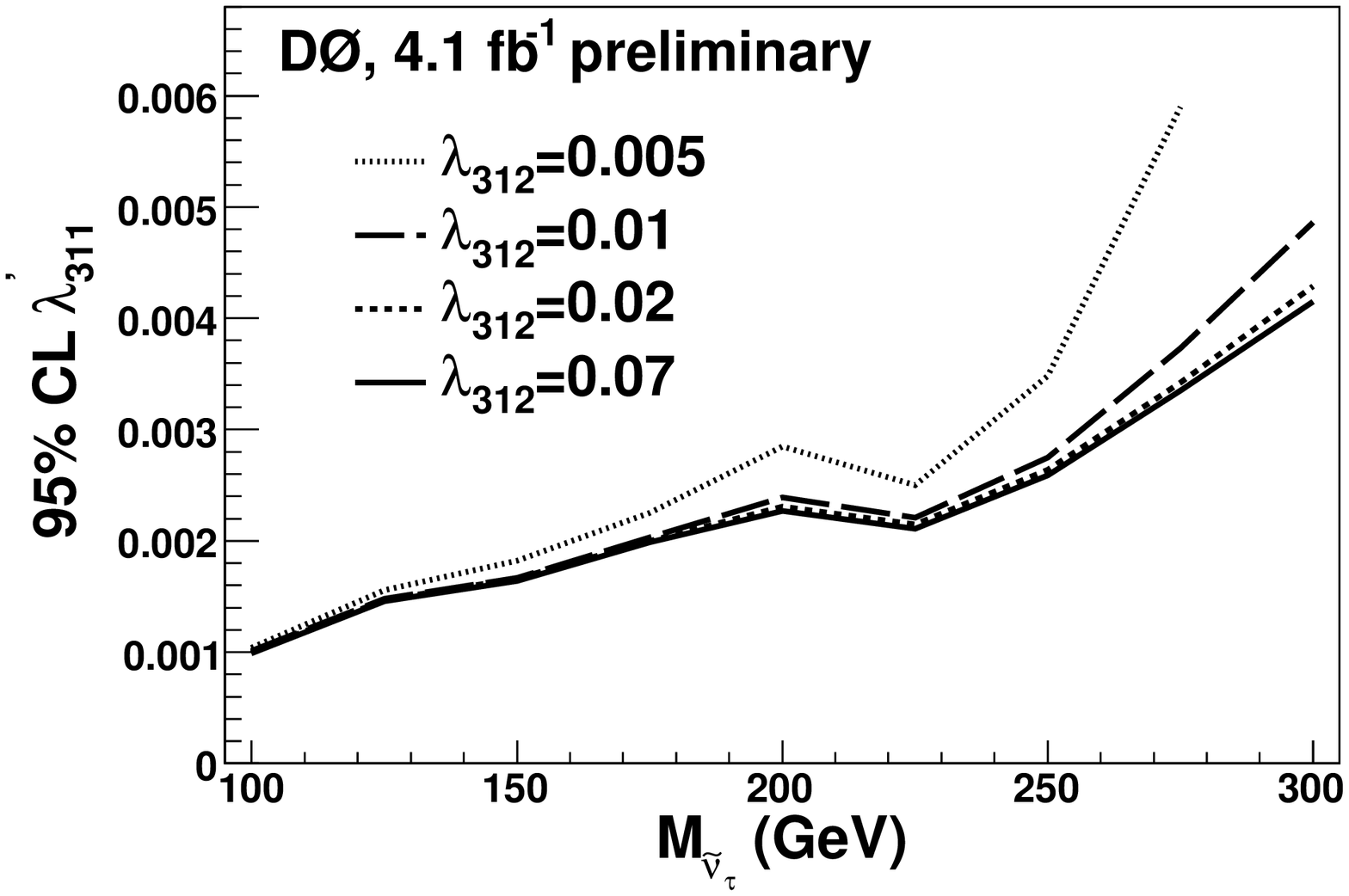}
    }
    \put(54,34){(a)}
    \put(120,34){(b)}
  \end{picture}
  \caption{(a) Distribution of the electron muon invariant mass, for data (points)
and background simulation (histogram) and expected signal of a 100~GeV sneutrino (dashed line).
(b) 95\% C.L. limits on RPV couplings as a function of the sneutrino mass.}
  \label{fig:d0_emu}
\end{minipage}
\end{figure}

Fig.~\ref{fig:d0_emu}a shows the $e\mu$ invariant mass distribution.
No indication of any new signal is observed as there is a good agreement
between data and the expected background.
Combining this new result with the published one, D0 provides on the total
sample a 95\% C.L. limit on the cross section times branching ratio 
(Fig.~\ref{fig:d0_emu}b).  
CDF has analysed a smaller data sample but has extended the analysis to $e\tau$ and $\mu\tau$
final states finding no excess over the SM expectation~\cite{cdf-emu+etau+mutau}. 
%
%
\vspace{-8mm}
\section{Diboson resonance search}
CDF has searched for a WW or WZ resonance decaying into the same
final state: an isolated electron with $E_T$ > 30~GeV and missing $E_T$ > 30~GeV (neutrino)
from the W and 2 jets with $E_T$ > 30~GeV from the second boson.
In addition, the events are required to have $H_T$ > 150~GeV where $H_T$ is the
scalar sum of the $E_T$ of the electron, neutrino and all jets with uncorrected $E_T$ > 8~GeV.   
The analysis~\cite{cdf-ww+wz} is based on an integrated luminosity of 2.9~fb$^{-1}$.
First the electron and missing $E_T$ are combined to form a W, generally there
are 2 solutions because of the unknown longitudinal component of the neutrino,
and both are kept otherwise the event is rejected.
The jet pair whose invariant mass is closest to the W (resp. Z) boson mass is used
to form the second boson if its mass is within the 65-105~GeV (resp. 70-100~GeV) range.

\begin{figure}[htb]
\centering
\begin{minipage}{0.9\textwidth}
  \centering
  \begin{picture}(145,40)
    \put(5,0){
      \includegraphics[width=6cm] {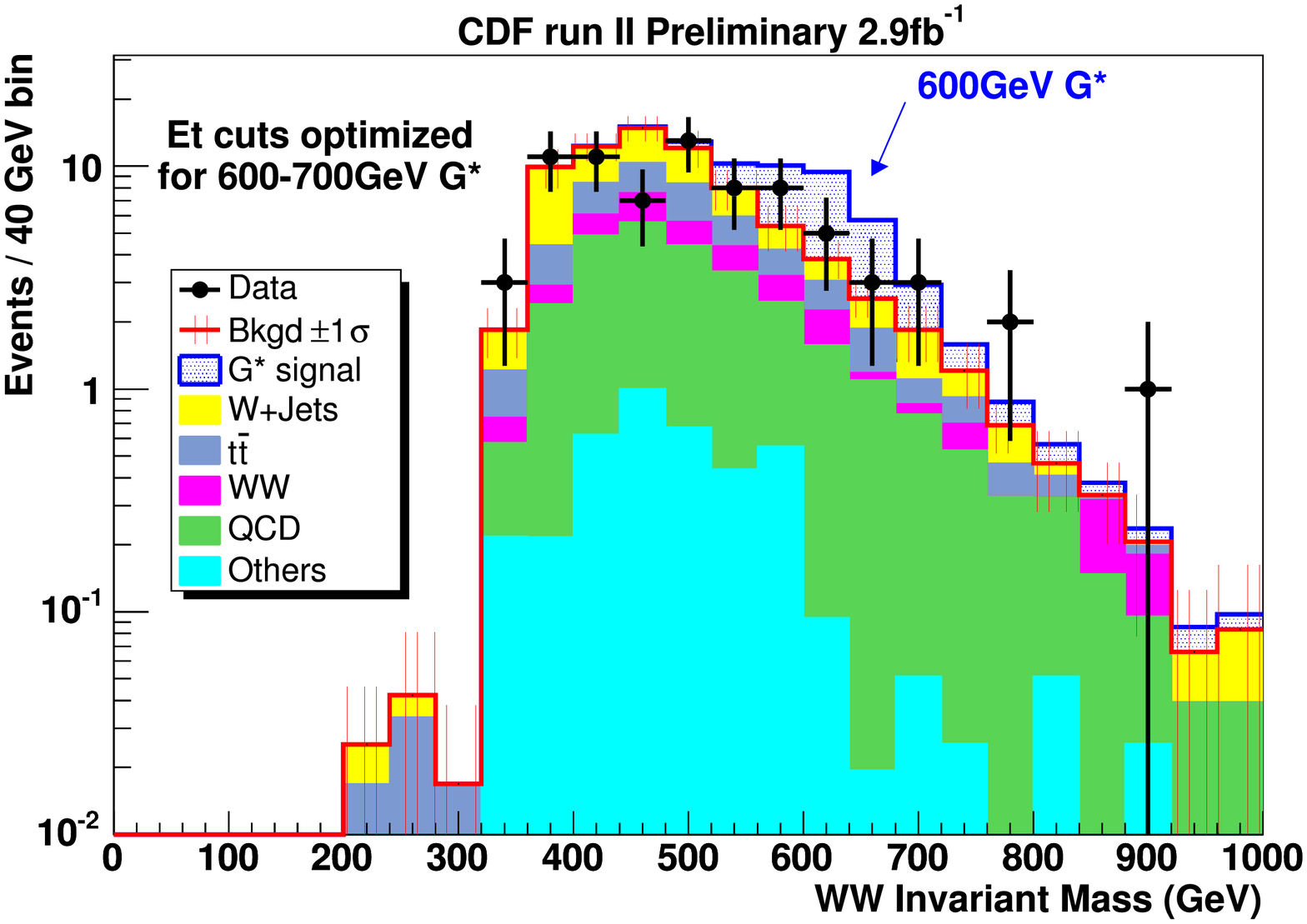}
    }
    \put(70,0){
      \includegraphics[width=6cm] {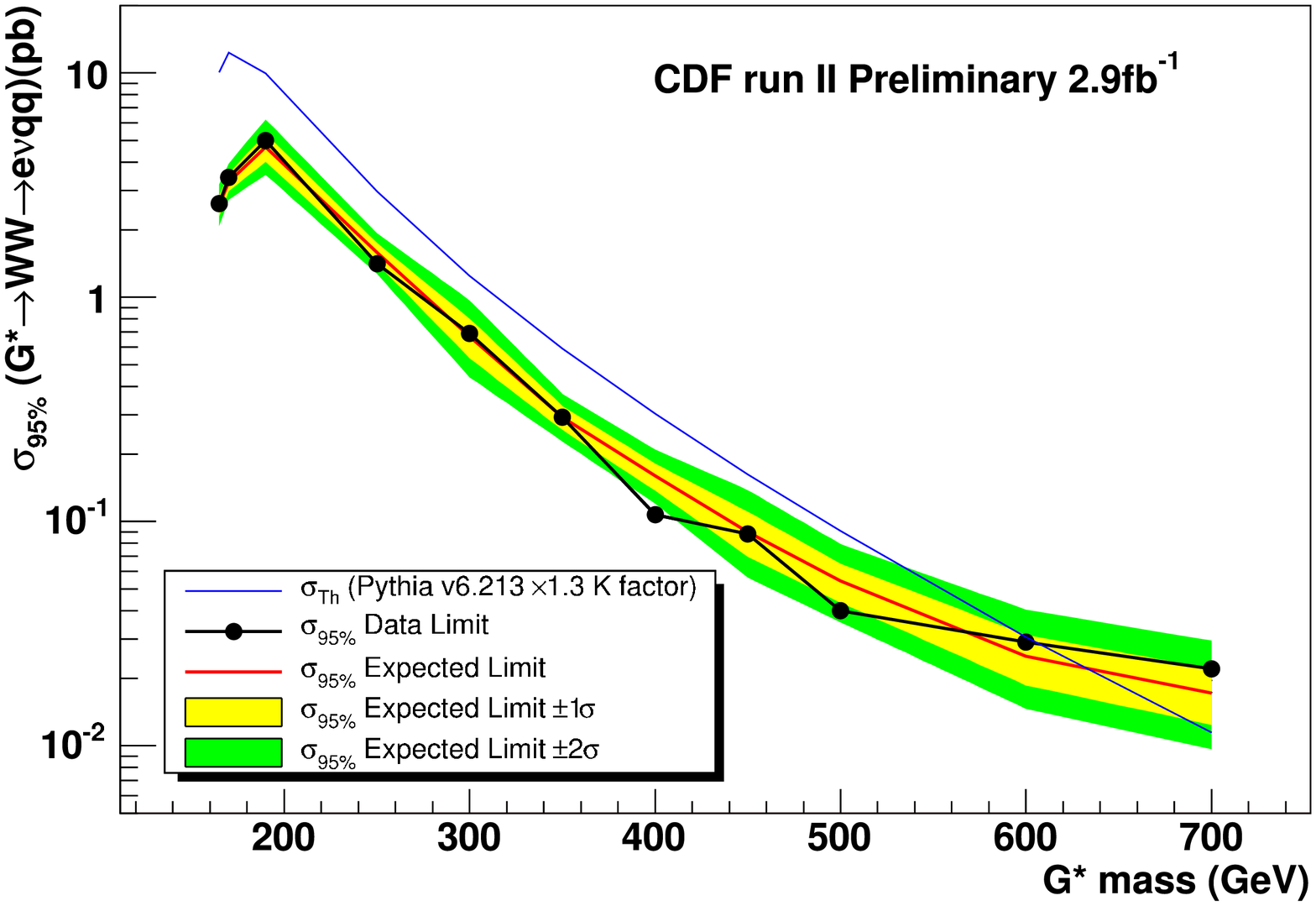}
    }
    \put(51,29){(a)}
    \put(116,29){(b)}
  \end{picture}
  \caption{(a) WW mass distribution after optimisation of $E_T$ cuts; the expected signal
of a 600~GeV graviton is added on top of the background;
(b) 95\% C.L. cross section limit for a graviton ($k/M_p$=0.1).}
  \label{fig:cdf_dibosons}
\end{minipage}
\end{figure}

Sensitivity to a new signal have been optimised for each mass by varying 
the $E_T$ cuts to get the best expected cross section limit at this mass.
As an example the data WW invariant mass distribution is compared with the
sum of all background contributions in Fig.~\ref{fig:cdf_dibosons}a after
this optimisation.
The dominating background is from W + jets events.
Contribution from an hypothetical 600~GeV graviton is added on top of the
background.
Fig.~\ref{fig:cdf_dibosons}b shows the 95\% C.L. cross section limit which is
obtained for a RS graviton with $k/M_{Pl}$=0.1.
%
%
\vspace{-8mm}
\section{Dijet resonance search}
CDF has measured the dijet differential cross section; the analysis based on a luminosity 
of about 1.1~fb$^{-1}$ is published~\cite{cdf-dijet}.
The dijet mass spectrum can be fitted by a smooth parametrisation over 7 orders of magnitude;
it shows no evidence for the existence of a new massive particle.
Limits on new particle production cross sections can be derived as a function
of the dijet mass.
They are translated into mass exclusion limits for a variety of models, although less
restricting than those obtained with lepton pairs.

\section*{Conclusion and acknowledgments}
Search for new resonances are still fruitless.
As the Tevatron collider is continuing to provide experiments with more data to analyze,
the quest for indications of new physics will be pursued by CDF and D0.

The author would like to thank the CDF and D0 working groups for providing
the material for this talk.
%

\end{document}